\documentclass[final,5p,times,twocolumn]{elsarticle}

\usepackage{hyperref,subcaption}

\begin{document}

\begin{frontmatter}

\title{Radionuclide Production for Dose Verification in VHEE FLASH Radiotherapy}

\author[1]{Francesco Urso}
\author[3]{Federica Baffigi}
\author[1,2]{Esther Ciarrocchi}
\author[3]{Leonida Antonio Gizzi}
\author[3]{Petra Koester}
\author[3]{Luca Labate}
\author[1,2]{Matteo Morrocchi}
\author[3]{Simona Piccinini}
\author[3]{Martina Salvadori}
\author[1,2]{Maria Giuseppina Bisogni}

\address[1]{Università di Pisa, Dipartimento di Fisica, Largo Pontecorvo 3, 56127 Pisa, Italy}
\address[2]{INFN, Sezione di Pisa, Largo Bruno Pontecorvo 3, 56127 Pisa, Italy}
\address[3]{CNR-INO, Istituto Nazionale di Ottica, Via Moruzzi 1, 56124 Pisa, Italy}

\begin{abstract}
This study explores the production of radionuclides in a PMMA (Polymethylmethacrylate) phantom irradiated with Very High Energy Electrons (VHEE) beams as a novel method for dose verification. In this work, Monte Carlo simulation studies using the Geant4 toolkit and experimental measurements with a VHEE beam produced by a laser-plasma accelerator were conducted. Specifically, the photonuclear production of $^{11}$C, $^{15}$O and $^{13}$N is examined, with a detailed analysis of their spatial distribution, production yields and decay time.\end{abstract}


\end{frontmatter}


\section{Introduction}
Radiotherapy (RT) is one of the main pillars of cancer treatment, using ionizing radiation to target and destroy malignant cells \cite{lievens2020provision,borras2016many}. While conventional modalities have demonstrated considerable efficacy, unintended damage to healthy tissues remains a major concern.

Recent advancements in this field have introduced FLASH radiotherapy, a method using ultra-high dose rate irradiation \cite{favaudon2014ultrahigh}.
FLASH RT leverages ultra-high dose rate (UHDR) irradiations-dose rates of $\geq$40 Gy/s delivered in less than 200 ms-compared to conventional dose rates ($\leq$0.03 Gy/s). This dramatic increase in dose rate has been shown to significantly spare normal tissues while maintaining tumor control \cite{wilson2020ultra,favaudon2014ultrahigh}. The potential benefits of FLASH RT, supported by promising preclinical data \cite{wilson2020ultra}, are driving ongoing efforts to translate this approach into clinical practice.

Very High Energy Electrons (VHEE), with energies in the range of 50 to 250 MeV, have emerged as a viable option for  the treatment of deep-seated tumors \cite{desrosiers2000150,lagzda2020influence,ronga2021back}. Unlike conventional electron beams (6--15 MeV), VHEE beams offer superior penetration and reduced sensitivity to tissue heterogeneities \cite{lagzda2020influence}, making them particularly suited for FLASH applications. Despite these advantages, the clinical implementation of VHEE FLASH RT faces several challenges, including the need for compact and cost-effective accelerators \cite{sarti2021deep,ronga2021back,faillace2022perspectives} and accurate dosimetry \cite{clements2024mini, hart2024plastic} .

In vivo verification of dose delivery is critical for ensuring that the administered dose matches the treatment plan. Although sophisticated monitoring systems have been established in proton therapy, similar real-time techniques are still lacking for VHEE radiotherapy. In proton and ion beam therapies, advanced monitoring methods such as Positron Emission Tomography (PET)\cite{parodi2000potential} and Prompt Gamma Imaging (PGI) \cite{wronska2020prompt} have been developed to verify dose delivery and range accuracy. For instance, PET imaging is based on the production of positron-emitting radionuclides such as $^{11}$C, $^{15}$O, and $^{13}$N generated during proton interactions with tissues. The subsequent detection of 511 keV annihilation photons enables three-dimensional mapping of the radionuclide distribution, thereby providing an indirect measure of dose deposition \cite{ptp_2020, bisogni2016inside}.

The production of radionuclides via photonuclear reactions during VHEE interactions presents an opportunity to adapt these well-established PET methods from hadron therapy to VHEE FLASH RT. When VHEE beams interact with a medium, high-energy gamma rays generated in the electromagnetic shower can induce photonuclear reactions that yield positron-emitting isotopes, primarily $^{11}$C and $^{15}$O. The spatial distribution of these radionuclides, once obtained with PET scanner, can be correlated with the dose distribution.

In this study, we initially investigated radionuclide production in a PMMA phantom irradiated with electrons in the VHEE energy range using Monte Carlo Geant4 simulations. These findings were subsequently validated through experimental measurements conducted with a laser-plasma accelerator system.

\section{Materials and Methods}
\subsection{Physics of Photonuclear Reactions}
When the VHEE beam interacts with matter, the electrons lose energy both through inelastic collisions and radiative emissions, the latter leading to the production of high-energy photons. These photons then convert into electron-positron pairs via pair production and the process continues, creating an electromagnetic shower until the energy of the electrons and positrons drops below the critical energy \cite{Leo_1994}.
High energy gamma rays generated in the electromagnetic shower interact with the nuclei of the medium via photonuclear reactions. The cross sections for these reactions are energy-dependent \cite{ishkhanov2004photonuclear}. The radionuclides produced in a PMMA phantom with their half-lives are listed in Table \ref{tab:radionuclides}. 
\begin{table}[htbp]
    \centering
    \caption{Radionuclides of interest in VHEE FLASH radiotherapy}
    \begin{tabular}{lll}
    \hline
    \textbf{Isotope} & \textbf{Production mechanism} & \textbf{Half-life (minutes)} \\
    \hline
    $^{11}$C & $^{12}$C$(\gamma, n)$$^{11}$C & 20.3  \\
    $^{15}$O & $^{16}$O$(\gamma, n)$$^{12}$O & 2.0  \\
    $^{13}$N & $^{14}$C$(\gamma, n)$$^{13}$N & 9.97  \\
    \hline
    \end{tabular}
    \label{tab:radionuclides}
\end{table}

\subsection{Simulation Setup}
Simulations were performed using the Geant4 toolkit \cite{Agostinelli2003} to model a PMMA phantom irradiated by a VHEE beam. The simulation framework incorporates detailed physics lists for electromagnetic interactions as well as photonuclear processes, modelled by the G4HadronPhysicsQGSP$\textunderscore$BIC$\textunderscore$HP physics list. 
Two simulation scenarios were considered: one with a full-size cylindrical phantom to study the development of the electromagnetic shower and another with a smaller box phantom to reproduce the experimental setup and validate the simulations.

The first simulated phantom had a radius of 6 cm and a length of 30 cm. This specific size was chosen to ensure proper development of the shower and to observe the activity as it varies with depth.

The electron beam was incident on the phantom along the central axis, with a Gaussian energy distribution centered at 150 MeV and a standard deviation of 1 MeV. The simulation was run for $10^8$ events to ensure statistical precision and the results scaled for the actual electron fluence of a FLASH irradiation.

\begin{figure}[htbp]
    \centering
    \includegraphics[width=0.5\textwidth]{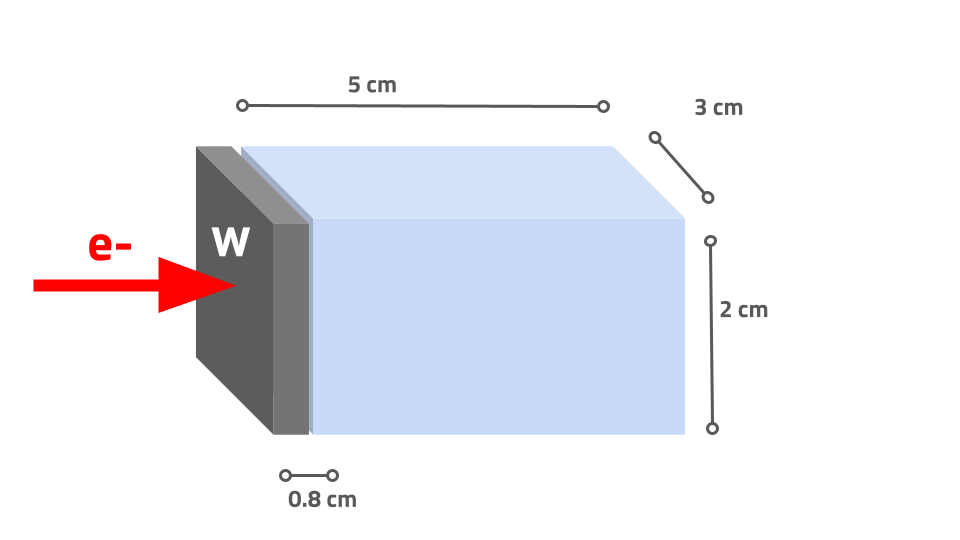}
    \caption{Schematic representation of the box phantom showing the tungsten converter sheet placed before a PMMA block to enhance photonuclear reactions.}
    \label{fig:schema}
\end{figure}

\begin{figure}[htb]
    \centering
    \includegraphics[width=0.45\textwidth]{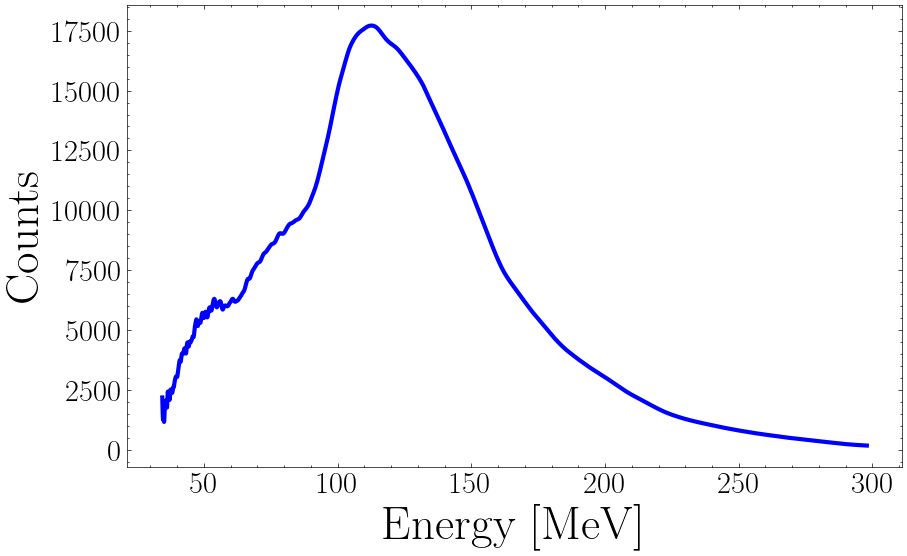}
    \caption{Energy spectrum of electrons generated by the laser-plasma accelerator at CNR-INO used for the experimental validation. This measured spectrum was also implemented in the simulation.}
    \label{fig:spectrum}
\end{figure}

For direct comparison with the experimental setup, a second set of simulations was carried out using the same configuration as in the experiment. In this simulation, a tungsten sheet, 0.8 cm thick, was placed immediately in front of the smaller box phantom (3x2x5 cm$^3$) to boost the generation of secondary gamma rays, thereby enhancing photonuclear reactions. The structure of the phantom is depicted in Figure \ref{fig:schema}. This setup, while not emphasizing the full development of the electromagnetic shower, provided a closer approximation to the experimental measurements. For this simulation the electron spectrum used was the one measured in the experiment. The spectrum was characterized using a magnetic spectrometer and is shown in Figure \ref{fig:spectrum}.

\subsection{Experimental Setup}
\begin{figure}[htb]
    \centering
    \includegraphics[width=0.35\textwidth]{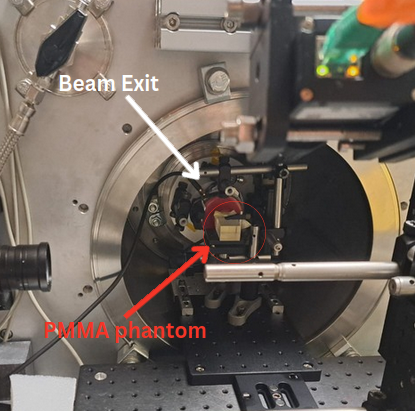}
    \caption{Laser-plasma accelerator experimental setup at CNR-INO showing the beam delivery system and the target for radionuclide production measurements.}
    \label{fig:setup}
\end{figure}

\begin{figure}[htb]
    \centering
    \includegraphics[width=0.35\textwidth]{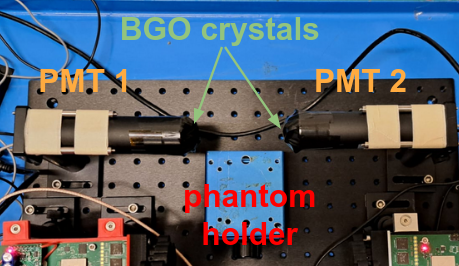}
    \caption{Coincidence detection system consisting of two BGO scintillation crystals coupled to Hamamatsu R9800 PMTs for measuring 511 keV annihilation photons.}
    \label{fig:photopmt}
\end{figure}

Experiments were performed at the National Research Council - Institute of Optics (CNR-INO) in Pisa using a laser-plasma accelerator\cite{labate2020toward}. The experiment was carried out using the box phantom with a tungsten sheet placed in front of it as described above. This choice was dictated by the limited space allowed in the experimental area. The experimental setup is shown in Figure \ref{fig:setup}.
After irradiation, the phantom was placed between two detectors operating in coincidence mode to detect annihilation events from radionuclides formed within the phantom. Each detector consisted of a BGO crystal coupled to a Hamamatsu R9800 PMT. The PMTs were connected to a Lecroy HDO 9204-MS oscilloscope, which recorded signals in coincidence with a 10 ns time window to select events. The coincidence detection system is shown in Figure \ref{fig:photopmt}.

The phantom was irradiated for 20 minutes with an electron beam containing approximately $10^{8}$ electrons per pulse at a repetition rate of 0.1 Hz.

\
\section{Results}
\subsection{Simulated Radionuclide Production and Distribution}

Simulation results presented here pertain to the full-size cylindrical phantom. Considering the conditions required for FLASH irradiation (dose rates $\ge$ 40 Gy/s), a pulse containing $10^{11}$ electrons was used as a reference for all simulations of the radionuclide production. Our results for the long cylindrical phantom show that under these conditions, the overall production reach $10^6$ radionuclides per pulse throughout the phantom, corresponding to an estimated induced activity of approximately $10^5$ Bq. Table \ref{tab:isotopes} reports, for each relevant nuclide, the number of nuclei produced per incident electron, specifically $^{11}$C, $^{13}$N, and $^{15}$O.

\begin{table}[htb]
    \centering
    \caption{Isotope production }
    \begin{tabular}{l r}
    \hline
    Isotope & Number of radionuclides/electron \\
    \hline
    $^{11}$C & 2x10$^{-6}$ \\
    $^{13}$N & 7x10$^{-7}$ \\
    $^{15}$O & 5x10$^{-6}$ \\
    \hline
    \end{tabular}
    \label{tab:isotopes}
    \end{table}

\begin{figure}[htb]
    \centering
    \includegraphics[width=0.4\textwidth]{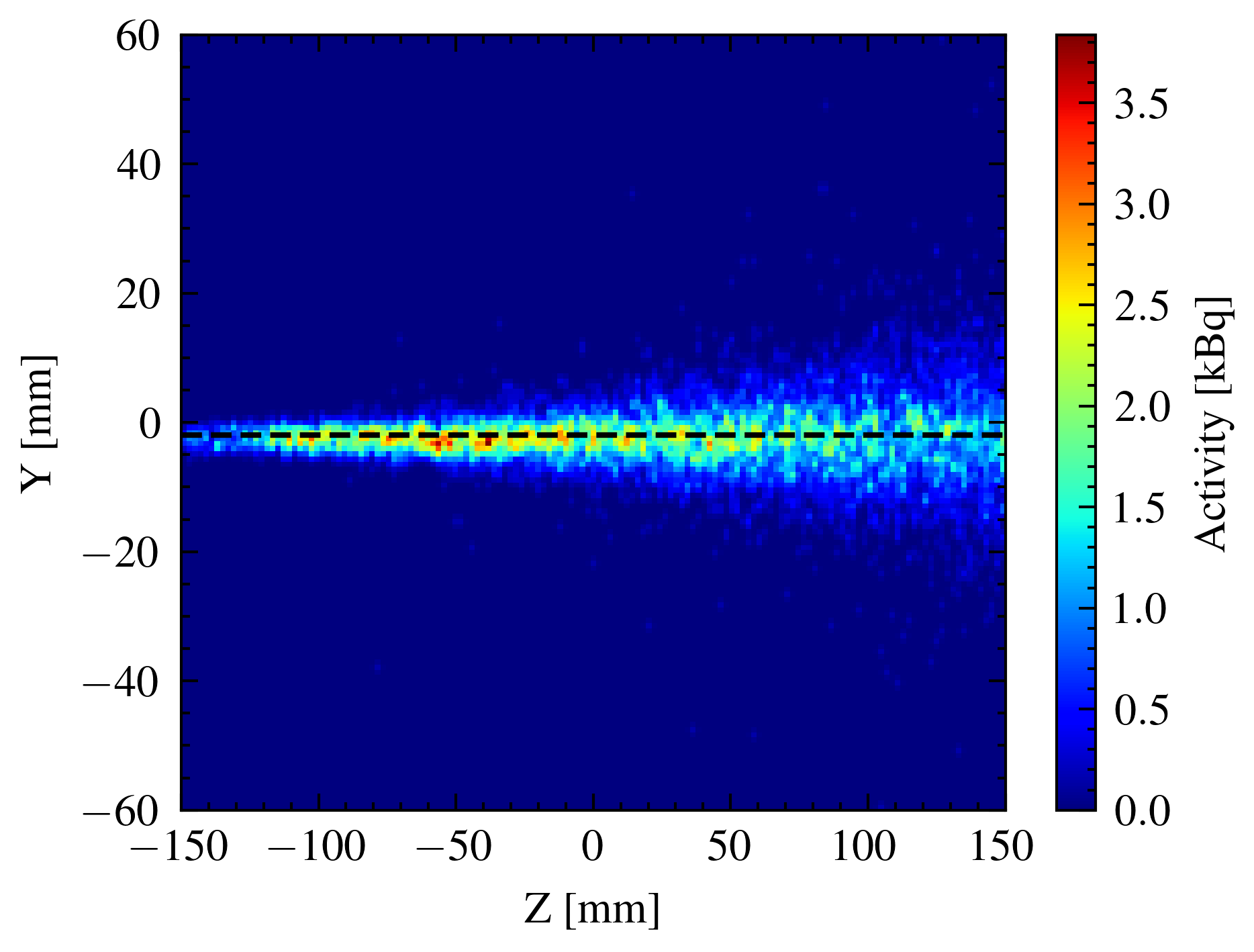}
    \caption{2D Spatial distribution of simulated $^{11}$C production in the full-size cylindrical PMMA phantom.}
    \label{fig:C11long}
\end{figure}

\begin{figure}
    \centering
    \includegraphics[width=0.4\textwidth]{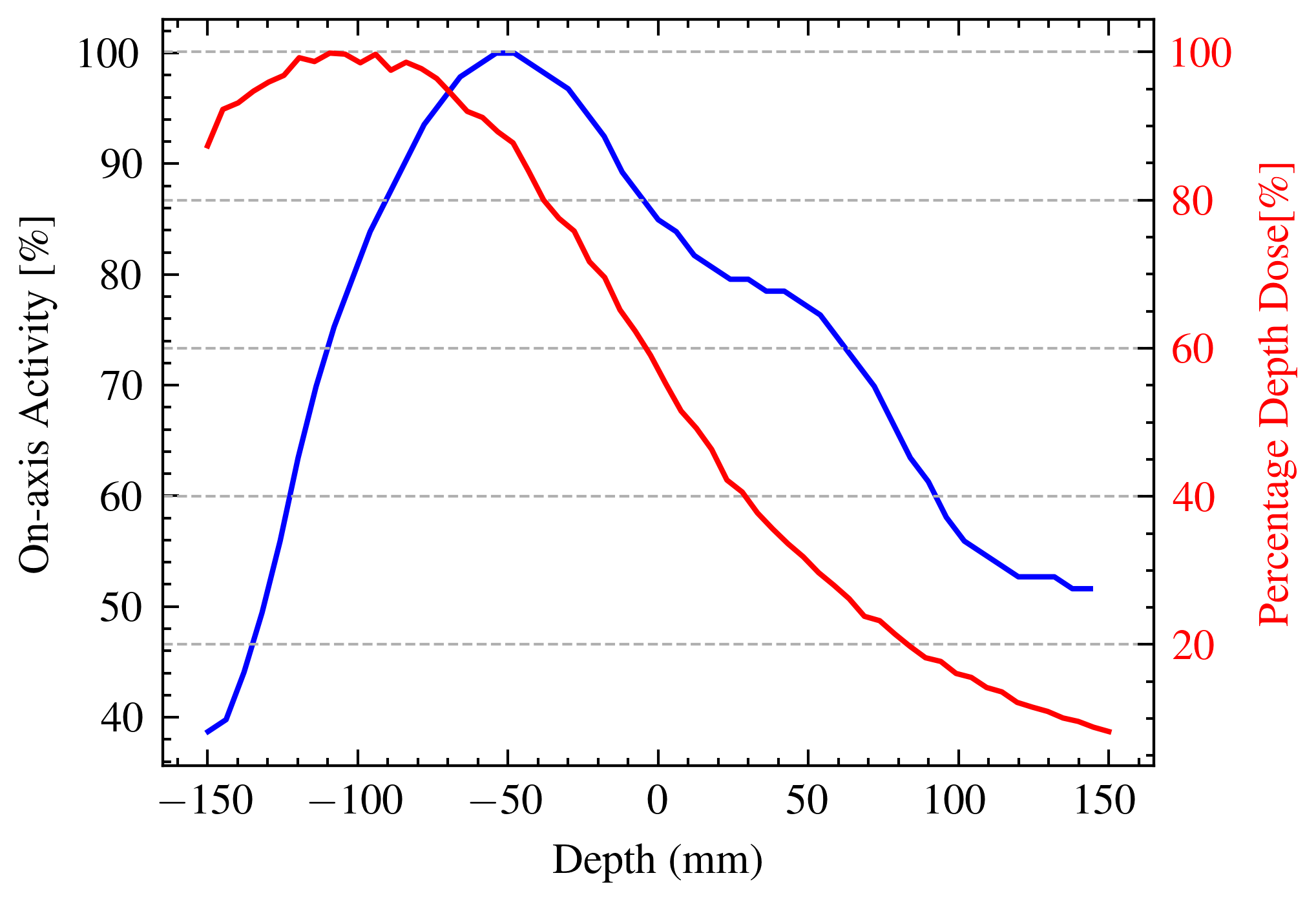}
    \caption{Activity profile (blue line) of $^{11}$C as a function of depth in the full-size cylindrical PMMA phantom and the dose distribution (red line) as a function of depth. The activity is normalized to the maximum value.}
    \label{fig:dosevc11}
\end{figure}

Figure ~\ref{fig:C11long} shows the 2D spatial distribution of the $^{11}$C activity produced in the PMMA cylindrical phantom in the longitudinal plane parallel to the cylinder  axis. The electron beam impinges on the phantom's entrance surface from the left side. Figure ~\ref{fig:dosevc11} shows the activity profile of $^{11}$C and the percentage depth dose along the central axis as a function of the depth in the PMMA phantom.

\subsection{Experimental Validation}

Experimental results and corresponding simulation validation were performed exclusively using the smaller box phantom setup. The primary focus of the experimental test was the detection of $^{11}$C, due to its favorable half-life for PET imaging. After irradiation, the box phantom was transferred to the detector described in Section 2. 

\begin{figure}[htb]
    \centering
    \includegraphics[width=0.4\textwidth]{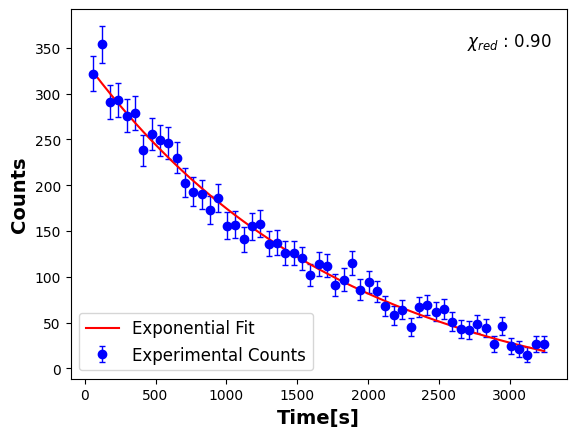}
    \caption{Coincidence counts from positron annihilation events measured at CNR-INO for the smaller box phantom. The exponential fit (red line) yields a half-life of $(20.9\pm0.2)$ minutes.}
    \label{fig:plasma}
\end{figure}

Figure \ref{fig:plasma} shows the exponential fit (red line) to the coincidence counts (blue dots), the fit resulted in a mean lifetime of $(30.2\pm0.2)$ minutes, corresponding to a half-life of $(20.9\pm0.2)$ minutes. These results are in  agreement with the expected half-life and confirm the production of $^{11}$C via photonuclear reactions in the PMMA phantom with the used electron fluence.

\begin{figure}[htb]
    \centering
    \includegraphics[width=0.4\textwidth]{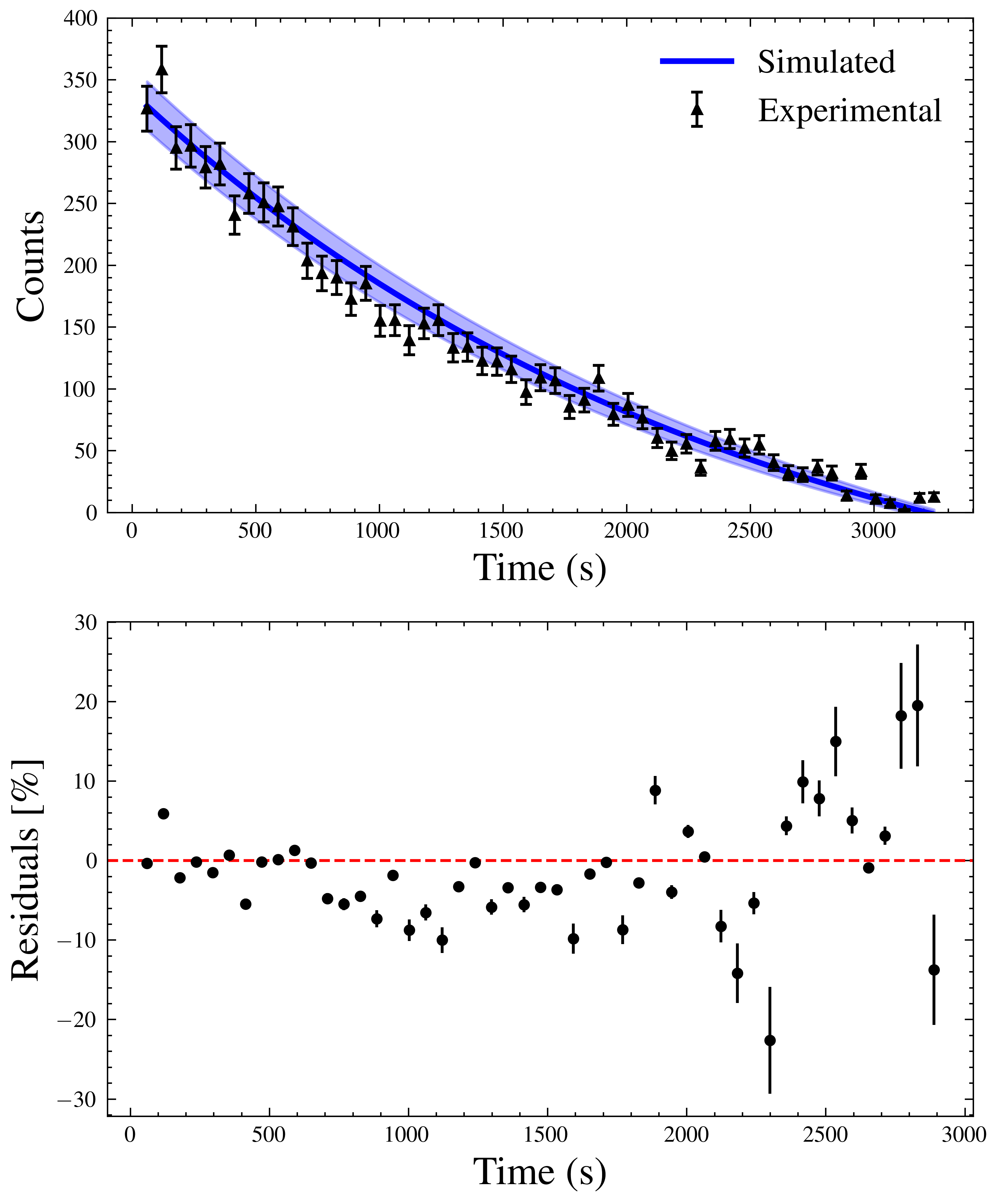}
    \caption{Comparison of simulated and experimental coincidence counts obtained from the smaller box phantom setup. The upper panel shows experimental and simulated counts with statistical uncertainties, while the lower panel displays the residuals.}
    \label{fig:plasmac11}
\end{figure}

Figure \ref{fig:plasmac11} presents a quantitative comparison between the experimentally measured and simulated $^{11}$C coincidence counts as a function of time. The upper panel displays the coincidence counts in black with the associated uncertainties, the blue line displays the simulated counts toghether with the interval of confidence (blue shaded area). 
The lower panel shows the residuals expressed as a percentage of the experimental counts. 

\subsection{Discussion}
Our simulations predict the production of positron-emitting isotopes under FLASH conditions, with approximately $2\times10^5$ $^{11}$C, $7\times10^4$ $^{13}$N, and $5\times10^5$ $^{15}$O nuclei per $10^{11}$ electrons. This activity levels, while modest compared to the activity administered in a nuclear medicine diagnostic exam, could be detectable with modern PET systems \cite{singh2024review}.

The spatial distribution of radionuclides in the cylindrical phantom shows distinct production patterns that do not correlate directly with dose. This discrepancy can be attributed to the fact that the photonuclear reactions are mediated by high-energy gamma rays produced in the electromagnetic shower, while the local dose deposition is mainly due to secondary electrons produced in the ionization processes.

The simulation results indicate that the maximum production of $^{11}$C occurs at depths that do not precisely align with the maximum dose deposition. This suggests that more sophisticated modeling is required to establish a correlation between produced activity and dose deposition.

The experimental verification confirms $^{11}$C production with a measured half-life (20.9±0.2 minutes) consistent with established values, validating our simulation framework. However, the comparison between experimental and simulated production reveals residual discrepancies that may stem from unaccounted systematic errors in the experimental setup.

The relatively short half-lives of the produced isotopes, particularly $^{15}$O ($\sim$2 minutes), imposes practical limitations on imaging, while $^{11}$C's longer half-life (20.3 minutes) allows more flexibility. This timing constraint could complicate clinical workflow, especially considering patient positioning and transport to imaging facilities.

Additionally, the current experimental setup using a tungsten converter enhances radionuclide production but differs from a realistic clinical configuration. Future work must bridge this gap by evaluating production under more clinically relevant conditions.
\section{Conclusions}
This study demonstrates the feasibility of detection of radionuclides produced during VHEE FLASH irradiation. Our Geant4 simulations predicted the production of positron-emitting isotopes ($^{11}$C, $^{15}$O, and $^{13}$N) in clinically relevant quantities, while experimental measurements confirmed $^{11}$C production with a measured half-life of $(20.9\pm0.2)$ minutes. The spatial distribution of these radionuclides provides valuable information about the beam interactions with the tissues, though the relationship between radionuclide density and dose remains complex.

Despite challenges including modest yield compared to conventional PET tracers and timing constraints due to isotope half-lives, our findings suggest that PET imaging of these radionuclides represents a possible approach for dose verification in VHEE FLASH radiotherapy. Future work should focus on optimizing detection protocols, refining the correlation models between radionuclide distribution and dose deposition, and evaluating performance under clinical treatment conditions.

\section*{Acknowledgments}
We acknowledge financial support under the Piano Nazionale Ripresa e Resilineza (PNRR), Mission 4, Component 2, Investment 1.1, Call for tender No. 104 published on 2.2.2022 by the Italian Ministry of University and Research (MUR), funded by the European Union – NextGenerationEU– Project Title Monitor for flash therapy (MORSE) – CUP I53D23000830006- Grant Assignment Decree No.  n. 974 adopted on 30/06/2023   by the Italian Ministry of Ministry of University and Research (MUR).

This work was also partially supported by the Piano Nazionale di Ripresa e Resilienza (PNRR), Missione 4, Componente 2, Ecosistemi dell’Innovazione – Tuscany Health Ecosystem (THE), Spoke 1 “Advanced Radiotherapies and Diagnostics in Oncology” - CUP I53C22000780001.

\bibliography{mybibfile}

\end{document}